\documentclass{jpp}
\usepackage{graphicx}

\usepackage[utf8]{inputenc}
\usepackage[T1]{fontenc}
\usepackage{amsmath}
\usepackage[numbers,sort&compress]{natbib}

\shorttitle{Anomalous coupling in radiation mediated shocks}
\shortauthor{A. Levinson, A. Granot, A. Vanthieghem and J. Mahlmann}

\title{Anomalous coupling in radiation mediated shocks}

\author{Amir Levinson\aff{1} 
  \corresp{\email{Levinson@tauex.tau.ac.il}}, Alon Granot \aff{1}, Arno Vanthieghem \aff{2,3}, Jens Mahlmann \aff{3}
}

\affiliation{\aff{1}School of physics and Astronomy, Tel Aviv University, Tel Aviv
\aff{2} International Research Collaboration Center, National Institutes of Natural Sciences, Tokyo 105-0001, Japan
\aff{3} Department of Astrophysical Sciences, Peyton Hall, Princeton University, Princeton, NJ 08544, USA
}

\begin{document}

\maketitle

\begin{abstract}
We summarize recent attempts to unravel the role of plasma kinetic effects in radiation mediated shocks.
Such shocks form in all strong stellar explosions and are responsible for the early electromagnetic emission
released from these events.  A key issue that has been overlooked in all previous works is the nature of 
the coupling between the charged leptons, that mediate the radiation force, and the ions, which are the dominant 
carriers of the shock energy.  Our preliminary investigation indicates that 
in the case of relativistic shocks, as well as Newtonian shocks in multi-ion plasma,
this coupling is driven by either, transverse magnetic fields of a sufficiently magnetized upstream medium, 
or plasma micro-turbulence if strong enough magnetic fields are absent.
We discuss the implications for the shock breakout signal, as well as abundance evolution and kilonova emission in binary neutron star mergers. 

\end{abstract}

\section{Introduction}
Astrophysical shocks are commonly divided into two main types: collisionless shocks, which form in optically thin regions, and are mediated by collective plasma instabilities on skin depth scales \citep{Blandford_1987,Spitkovsky_2005}, and radiation mediated shocks (RMS), which form in optically thick regions and are mediated by
Compton scattering and, under certain conditions, pair creation on scales of the order of the Thomson length 
(see \citealt{katz2017,levinson2020}, and references therein).  

Formation of a RMS occurs when,
(i) the photon diffusion time across the shock is comparable to the shock crossing time
of the flow\footnote{This is also the deceleration length of the flow. To see this note that the mean force acting on a baryon (assuming tight coupling between ions and electrons)
is $-m_p d\beta/dz = \sigma_T e_{rad}$, where $e_{rad}$ is the local energy density of the radiation. 
Energy conservation yields $e_\gamma \approx n_u m_p c^2 \beta_u^2$ in the immediate post shock region, where $\beta_u$ is the shock velocity and $n_u$ is the plasma density upstream, with which one obtains $-d\beta/dz\sim \beta_u/L_{dec} \sim \sigma_T n_u \beta_u^2$.  Thus, $L_{dec}\simeq (n_u\sigma_T\beta_u)^{-1}$ equals the photon diffusion length (see 
 \citealt{levinson2020}). A similar result can be derived for relativistic shocks \citep{granot2018,vanthieghem2022}.}, and  (ii) the shock is radiation dominated  \citep[e.g.,][]{weaver1976,katz2010}.   
Condition (i) requires the optical depth ahead of the shock to
exceed $\beta_u^{-1}$ (roughly the shock width), 
where $\beta_u$ is the shock velocity in units of $c$  
\footnote{In relativistic RMS, this scaling is altered by Klein-Nishina effects and excessive pair production.}. 
Condition (ii) requires the shock to be fast enough, $\beta_u > 2\times10^{-4} (n_u/10^{15}\, {\rm cm^{-3}})^{1/6}$, here $n_u$ is the upstream density.   
These two conditions pertain in essentially all strong stellar explosions, including 
various types of supernovae, low luminosity GRBs, regular GRBs and binary neutron star (BNS) mergers \citep{levinson2020}. 
Consequently, RMS are inherent features of stellar explosions. 
The RMS structure and dynamics depend on the progenitor type, the explosion energy, and the explosion geometry. In particular, 
the shock velocity at breakout can range from sub-relativistic to mildly relativistic, and in extreme cases even ultra-relativistic.

The prime motivation to explore the physics of RMS is the desire to predict the properties of the radiation emitted during the breakout 
of the shock from the opaque envelope surrounding the explosion center. This breakout emission -  the first electromagnetic signal 
a distant observer can detect - and the subsequent cooling emission, carry valuable information regarding the explosion mechanism and 
the progenitor type, which is otherwise inaccessible.  Moreover, as explained in \S \ref{sec:NRMS_ion_implc} 
(see \citealt{granot2023}  for a detailed analysis), in case of BNS mergers
the RMS can trigger nuclear transmutations prior to its breakout, that can significantly impact the abundance evolution of r-process material 
and the resultant kilonova emission.  

The physics of RMS involves two disparate scales; the Thomson scattering length,
 \begin{equation}
\lambda= (\sigma_T n_u)^{-1} \approx 10^9 \left(\frac{n_u}{10^{15}\, cm^{-3}}\right)^{-1} \quad \rm cm, \label{eq:lambda}
\end{equation}
that dictates the global shock dynamics and structure, and the (proton) skin depth,
 \begin{equation}
l_p = \frac{c}{\omega_p} \approx 0.5 \left(\frac{n_u}{10^{15}\, cm^{-3}}\right)^{-1/2} \quad \rm cm, \label{eq:skindepth}
\end{equation}
on which coupling between ions, electrons and, in relativistic RMS, positrons is anticipated to occur.
Here $\omega_p=(4\pi e^2 n_u/m_{\rm p})^{1/2}$ is the (proton) plasma frequency.
As seen, under typical astrophysical conditions the separation between kinetic and radiation scales is huge.
This is presumably the reason why the issue of ion-electron coupling in RMS has been overlooked until very recently. 
In what follows, we outline recent attempts to investigate the role of plasma kinetic effects in RMS, and discuss potential
implications for observations of stellar explosions. 

The interaction of intense radiation field with plasma has been studied also in other astrophysical contexts.  In particular,
the dynamics of Compton-driven plasma wakefields generated by the interaction of GRB prompt emission with
the circumburst medium has been studied in \citet{Fred2008} and \citet{guadio2020}.  A kinetic study 
on the development of plasma instabilities in bright GRBs is presented in \citet{martinez2021}.

\section{The role of plasma turbulence in the coupling of pairs and ions}\label{sec:plasma_turbulence}
All models previously developed to compute the structure and emission of RMS invoke the single-fluid approach, which 
implicitly assumes infinitely strong coupling between all plasma constituents.
However, since the cross section for Compton scattering off protons is smaller by a factor $(m_{\rm p}/m_{\rm e})^2$ than that for electrons
(and positrons when present), 
the radiation force acting on the ions is completely negligible.   This raises the question of how the radiation force is 
mediated to the ions.  The conventional wisdom has been that a tiny charge separation, induced by the radiation force experienced
by the charged leptons, generates electrostatic field that decelerates the ions, and apart from providing a mediator of the
radiation force this coupling has no further consequences for the shock properties beyond the single fluid models.  
Indeed, in case of a Newtonian RMS that propagate in a pure hydrogen gas, a velocity separation of
\begin{equation}
\Delta\beta/\beta_u  \approx (l_p/\lambda)^2 \lesssim 10^{-19}
\end{equation}
is sufficient to generate the required electric field that decelerates the protons  \citep{levinson2020b}. 
However, as shown below, in the presence of positrons and/or ion species with different charge-to-mass ratio electrostatic coupling fails, and
a different mechanism is needed to couple the various plasma constituents.

\subsection{Relativistic RMS}\label{sec:RRMS}
In sufficiently fast RMS ($\beta_{u} \gtrsim 0.5$) the temperature is high enough to allow rapid creation of 
$e^+e^-$  pairs inside the shock \citep{levinson2008,budnik2010,katz2010,ito2020,ito2020b}.  For $\gamma_{u}>1$ the pair multiplicity approaches $\gamma_{u} m_{\rm p}/m_{\rm e}$,
completely dominating the shock opacity.    Under such conditions electrostatic forces cannot provide coupling 
between pairs and ions since the electric field required to decelerate the ions exerts opposite forces on electrons and positrons. 
To illustrate this, a multi-fluid model for unmagnetized, relativistic RMS (RRMS) propagating in a pure hydrogen 
plasma has been developed recently \citep{levinson2020b}.  
The analysis indicates that once the density of newly created positrons 
 approaches the baryon density, which in RRMS occurs at the onset of the shock transition layer,  
 the charge density, and ultimately the electric field, reverse sign.
 This leads to decoupling of the different species early on, and to the development of relative drifts between the different beams. 
The presence of a strong enough background magnetic field perpendicular to the shock velocity can lead to tight coupling, 
preventing velocity spreads \citep{mahlmann2023}.

The velocity separation between the different plasma constituents imposed by the radiation force in unmagnetized RRMS, 
is expected to induce a rapid growth of plasma instabilities.
 Linear stability analysis \citep{vanthieghem2022} indicates a growth of various plasma modes, which ultimately become dominated
by a current filamentation instability driven by the relative drift between the ions and the pairs.
Particle-in-cell simulations \citep{vanthieghem2022} validate these results and further probe the nonlinear regime of the instabilities, 
elucidating the pair-ion coupling by the microturbulent electromagnetic field.  These simulations are local, in the sense that they encompass 
kinetic-scale region inside the shock transition layer.  The radiation force is modeled as a prescribed force acting solely on the pairs.  The 
relevancy of such simulations to realistic RMS stems from the huge scale separation between radiation and kinetic physics.  

An interesting result found in \citet{vanthieghem2022} is a dependence of the coupling length on the pair multiplicity 
$\mathcal{M}$ (it scales roughly as $\mathcal{M}^{1/2}$).  For the large multiplicity anticipated well inside the shock, $\mathcal{M}\approx \gamma_u m_{\rm p}/m_{\rm e}$,
it becomes macroscopic,  $\sim10^6 l_p$.  This is still much smaller than the width of an infinite, planar RRMS, but might exceed the shock 
width during the breakout phase (see \S \ref{sec:long_rang} for further discussion).

\begin{figure}
  \centering
  \includegraphics[width=1.\textwidth]{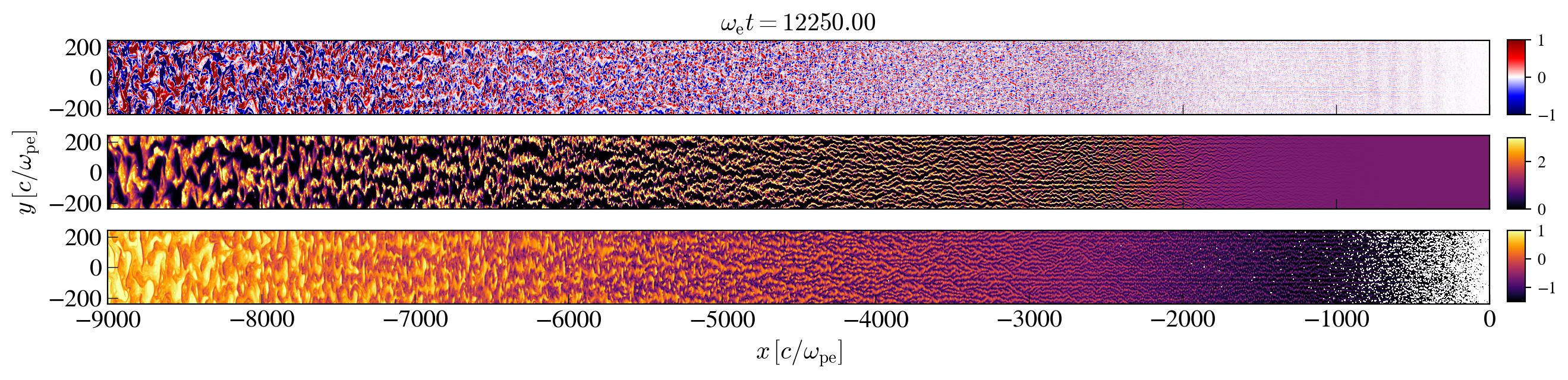}
  \caption{ Structure of the RMS precursor for fiducial parameters described in the text. We determine the position $x\,=\,0$ as the tip of the photon beam. (Top) Transverse magnetic field profile; (Middle) Density of the ions; (Bottom) Density profile of the injected positrons in log-scale. As seen, the combined lepton deceleration and pair loading lead to progressive current filamentation of the background, followed by anomalous coupling of the different species.}
     \label{fig:Test_RMS}
\end{figure}

While small-scale simulations allow us to capture more realistic values of the effective radiation force acting on leptons, they do not account for the kinetic and dynamical coupling between microturbulence, pair loading, and spatially varying radiation force. Here, we extend the previous hybrid description to investigate this effect, by accounting for the contribution of Compton scattering and pair production from an idealized photon distribution as described in Levinson (2020). Specifically, the photon density of the beam propagating away from the shock provides the magnitude of the lepton recoil from Compton scattering, while the density of each beam shapes the local injection rate. In this cold beam approximation, we do not resolve the energy-momentum tensor of the photon distribution, and hence, the energy of the pairs at injection is not constrained. As one would expect in a realistic configuration, we assume that the pairs are injected at rest into the local lepton frame with an ad hoc temperature. Below, we discuss a first attempt to perform global hybrid simulations of the precursor of a relativistic radiation-mediated shock with fluid photons and kinetic electrons, positrons, and ions.

Figure 1 shows the results of a hybrid simulation performed using the OSIRIS code \citep{10.1007/3-540-47789-6_36,2013PPCF...55l4011F}. In this example, plasma is injected into the right-hand side boundary with a Lorentz factor of $\gamma\,=\,10$, weak background magnetic field of $eB_z\,=\,0.1\,m_{\rm e} \omega_{\rm pe} c$, and reduced mass ratio of $m_{\rm i}\,=\,25\,m_{\rm e}$. We resolve the plasma skin depth $\Delta x \,=\, 0.7\,c/\omega_{\rm pe}$ where $\omega_{\rm pe}^2\,=\,4 \pi \gamma_\infty n_\infty e^2/m_{\rm e}$, $c\Delta t\,=\,0.5\,\Delta x$ corresponding to the CFL of electromagnetic field solver introduced in ~\cite{2018CoPhC.224..273B} and extensively tested in~\cite{2022ApJ...933...74G}.  The lepton recoil is artificially enhanced to observe significant deceleration over the box size. The combined lepton deceleration and pair loading lead to progressive current filamentation of the background, followed by anomalous coupling of the different species.


\subsection{Shocks propagating in multi-ion plasma}\label{sec:multi_ion_RMS}
In reality, the medium into which the shock propagates in most of the aforementioned systems contains multiple ion species with different 
charge-to-mass ratio.  In such situations, the RMS physics may be considerably altered.
The main point to note is that the deceleration rate of ions inside the shock depends on the charge-to-mass ratio, and since the charge conservation 
condition in a multi-ion plasma is degenerate, a large velocity separation 
between the different ions in the post-deceleration zone is, in principle, allowed, even in sub-relativistic RMS that are devoid of positrons. 
As in RRMS, sufficiently strong magnetic fields can couple the ions, however, 
since the gyroradius of an ion of mass $Am_{\rm p}$ is larger by a factor $Am_{\rm p}/Zm_{\rm e}$ than that of an electron (or positron), much stronger magnetic fields may be needed to couple all ions. Such strong fields may not be present in most relevant systems.

\begin{figure}
  \centering
  \includegraphics[width=0.4\textwidth]{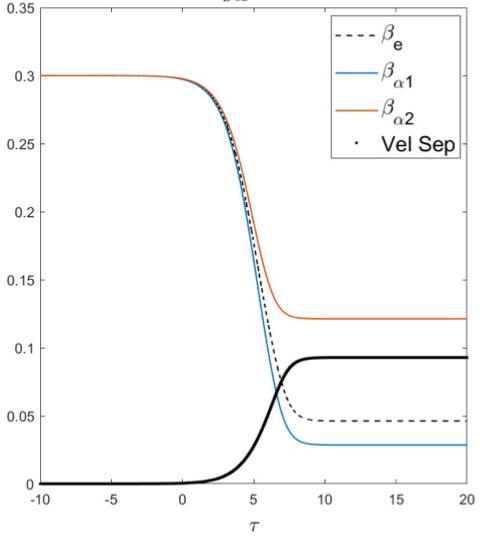}
  \caption{Velocity profiles of electrons (dashed line)  and two ion species, $\alpha1$ (solid red) and $\alpha2$ (solid blue) inside
    a shock moving at a velocity of $\beta_u=0.3$.  The charge-to-mass ratio in this example is 
    $Z_{\alpha_1}/A_{\alpha_1}=0.47$ and  $Z_{\alpha_2}/A_{\alpha_2}=0.4$,
 and the density ratio is $n_{\alpha_1}/n_{\alpha_2} = Z_{\alpha_2}/Z_{\alpha_1} $, where $Z_{\alpha_i}$ and $A_{\alpha_i}$ denote the atomic 
 and mass numbers of species $\alpha_i$, respectively.  The solid black line gives the velocity separation, $\beta_{\alpha_2} - \beta_{\alpha_1}$.}
     \label{fig:RMS_ion}
\end{figure}

The structure of a Newtonian, multi-ion RMS has been explored recently using a semi-analytic, multi-fluid shock model 
that can incorporate any number of ion species (Granot, Levinson \& Nakar, in preparation). 
The model invokes the diffusion approximation for the transfer of radiation through the shock, and computes the electrostatic coupling between the ions and electrons in a self-consistent manner, by solving the energy and momentum equations of the radiation and the multi-fluid plasma, together with
Maxwell's equations, taking into account the electrostatic force acting on the charged fluids.  An example is shown in Fig. \ref{fig:RMS_ion} for
two ion species.  As seen, a substantial velocity separation is developed inside the shock, and is maintained in the post deceleration zone,
where all forces (radiation and electrostatic) vanish.  In practice, the relative drift between the different ion beams (and the electrons) is expected to lead
to generation of plasma turbulence that will ultimately couple the ions, transferring their energy to radiation in the downstream.  
However, the dominant wave modes should differ from those found in a single-ion RRMS \citep{vanthieghem2022}.

In general, the anomalous coupling length is expected to depend on the threshold drift velocity for the
onset of the instability, on the saturation level of the turbulence, and conceivably other factors.   
In BNS mergers, where the plasma density $\gtrsim 10^{25}$ cm$^{-3}$, the anomalous coupling length may be large enough to allow full velocity separation,
leading to nuclear transmutations that might have important implications to kilonova emission (see \S \ref{sec:NRMS_ion_implc} for further discussion).  
In other systems, where the density is lower by many orders of magnitudes,  anomalous coupling likely occurs on 
scales much smaller that the radiation length. In that case, dissipation via anomalous friction should first lead to formation 
of collisionless subshocks  before the energy will be converted to radiation.  If the growth of the instability occurs well after decoupling, the 
ion temperature should be high enough to allow inelastic ion-ion collisions by random motions.  A full kinetic study is needed to compute the detailed shcok 
structure and assess whether nuclear reactions are expected in such systems.

\section{Implications for stellar explosions}\label{sec:implications}
In the following, we highlight potential implications of the results  hitherto discussed
for breakout dynamics and emission, as well as changes in composition.

\subsection{Generation of nonthermal particle distributions and hard emission}

The generation of magnetic turbulence in RRMS can potentially convert a fraction of the 
dissipated energy to nonthermal particles with power-law distributions. Both the
increased nonadiabatic heating found in \citet{vanthieghem2022} and the possible
formation of power-law energy spectra can impact the shock breakout emission. 
In particular, a hard spectral component, extending well beyond current predictions of single-fluid models,
may be present in the breakout signal.

\subsection{Long range plasma scales at high pair multiplicity}\label{sec:long_rang}
Single-fluid models of finite RRMS \citep{granot2018}, indicate that during the shock breakout 
phase the width of shock transition layer decreases dramatically, owing to radiative losses.
Since the scale over which pairs and ions couple becomes macroscopic when the pair multiplicity becomes large 
(see  \S \ref{sec:RRMS}), it could be that the coupling length will approach or even exceed the radiation scale during the shock breakout episode. 
If this indeed happens, it means that the RMS structure will be vastly different 
than predicted by current RRMS models, which might have profound implications for the observed shock breakout signal
(e.g., in the case of a shock breakout from a stellar surface a wider shock implies a softer and more energetic breakout radiation).
Under such circumstances, global shock models may be necessary to compute the breakout dynamics and emission.

\subsection{Abundance evolution and kilonova emission in BNS mergers}\label{sec:NRMS_ion_implc}
The expulsion of a relativistic jet following neutron star coalescence drives a fast RMS into the merger ejecta. 
The breakout of the shock from the high-velocity tail of the merger ejecta can produce a gamma-ray flash that, even though much fainter than 
a typical short GRB, can overwhelm the jet emission at large enough viewing angles (with respect to the jet axis).  According to one scenario \citep{kasliwal2017,gottlieb2018,pozanenko2018,beloborodov2020},
the gamma-ray flash GRB 170817A that accompanied the gravitational wave signal GW 170818 was produced by such a process.

As explained in \S \ref{sec:multi_ion_RMS}, the propagation of the RMS through the merger ejecta can lead to a significant velocity separation of 
different r-process isotopes just downstream of the shock transition layer, provided that anomalous friction is not too effective.  An estimate
of the scale separation in RMS  suggests that, under conditions anticipated in BNS mergers, the anomalous coupling length is likely to exceed the 
shock width (Granot, Levinson \& Nakar, in preparation).
In that case, collions of the different ion beams will induce nuclear transmutations in regions where the shock velocity exceeds the 
corresponding activation barriers.  Recent analysis (Granot, Levinson \& Nakar, in preparation) indicates that the ion-ion collision length
is smaller than the shock width, and that in regions where  $\beta_u \gtrsim 0.2$, the collision energy may be large enough
to induce fission and fusion of many elements.   This can significantly alter
 the composition profile of r-process material behind the shock and, potentially,
the kilonova emission if the change in composition affects the opacity and/or the radioactive energy deposition in the ejecta.  

In addition to inelastic ion collisions, neutron-rich isotopes downstream of the shock can undergo fission through the capture 
of free neutrons that cross the shock.   
The activation energy for neutron-induced fission ranges from practically zero for $^{235}U$ to a peak of about $40 MeV$ for elements of mass number $A\approx100$,
so a shock velocity of $\beta_u \gtrsim 0.1$ should give rise to fission of many elements. 
Whether free neutrons are sufficiently abundant at early times (during shock propagation) 
to significantly alter the ejecta composition is yet an open issue.  

\section{Summary} 

A fundamental question in the theory of radiation mediated shocks is: What is the mechanism
that couples the different plasma constituents (ions, electrons and positrons), and how does it affect the 
shock thermodynamics and emission?  We have shown, by means of semianalytic approach and 
PIC simulations, that this coupling is accomplished through generation of plasma turbulence in weakly magnetized shocks,
or by magnetic fields in sufficiently magnetized, perpendicular shocks.  The generation of plasma microturbulence in weakly magnetized
RMS can lead to particle acceleration inside the shock, which might  alter the characteristics of the shock breakout emission.
Moreover, our simulations indicate that the coupling length increases with increasing pair multiplicity, becoming macroscopic 
in regions where the pair multiplicity saturates.  
This could considerably affect the shock structure and emission during the breakout phase, e.g., due to early formation of collsionless subshocks,
since then the shock thickness is expected to shrink dramatically, owing to radiative losses, ultimately subceeding the coupling length.

We also find that in shocks propagating in multi-ion plasma, 
a substantial velocity separation between ions having different charge-to-mass ratio develops inside the shock.  
This can lead to inelastic nuclear collisions that can induce fission (or fusion).  In BNS mergers, 
the capture of free neutrons that cross the shock by neutron rich isotopes downstream of the shock can also induce fission.  
The passage of a fast enough shock in BNS merger ejecta can, therefore, lead to 
a considerable evolution of the relative abundance of r-process elements at early times, which could have important consequences for the kilonova emission.

\section*{acknowledgment}
We thank the referee for useful comments. AL and AG acknowledge support by the Israel Science Foundation grant 1995/21.
AV acknowledges support from the NSF grant AST-1814708 and the NIFS Collaboration Research Program (NIFS22KIST020).
JFM acknowledges support by the National Science Foundation under grant No. AST-1909458. This research was supported
by the Multimessenger Plasma Physics Center (MPPC), NSF grant PHY-2206607.

\section*{Declaration of Interests} 
The authors report no conflict of interest.

\bibliographystyle{jpp}

\bibliography{RMS}

\end{document}